\documentclass{article}

\usepackage{amssymb}
\newcommand{\be}{\begin{equation}}
\newcommand{\ee}{\end{equation}}
\newcommand{\bea}{\begin{eqnarray}}
\newcommand{\eea}{\end{eqnarray}}
\newcommand{\bean}{\begin{eqnarray*}}
\newcommand{\eean}{\end{eqnarray*}}
\title{Boundary conditions for the spinor field in Rindler spacetime
and the quantum field theoretical basis of the Unruh effect}
\author{Daniele Oriti\thanks{d.oriti@damtp.cam.ac.uk  -  oriti@icra.it} \\ Department of Applied Mathematics and
Theoretical Physics, \\ Centre for Mathematical Sciences, University of
Cambridge, \\ Wilberforce Road, Cambridge CB3 0WA, UK \\ and \\ St. Edmund`s
College, Cambridge CB3 0BN, UK \\ and \\ I.C.R.A.-International Center for
Relativistic Astrophysics,  \\ University of Rome ``La Sapienza'', \\ P.le
A.Moro, 5 - 00185 - Rome - Italy}

\begin{document}
\maketitle
\begin{abstract}
We analyse the quantization procedure of the spinor field in the
Rindler spacetime, showing the boundary conditions that should be
imposed to the field, in order to have a well posed theory. Because of
these boundary conditions we argue that  this construction and the usual
one in Minkowski spacetime are qualitatively different and can not be
compared and consequently the conventional interpretation of the Unruh
effect, that is the thermal nature of the Minkowski vacuum state from
the point of view of an accelerated observer, is questionable. We also
analyse in detail the Unruh quantization scheme and we show that it is
not valid in the whole Minkowski space but only in the double Rindler
wedge, and it cannot be used as a basis for a quantum theoretical proof
of the Unruh effect.
\end{abstract}
\section{Introduction}
The \lq\lq Unruh effect" could be expressed by the following statements: 
 1) the Minkowski vacuum state, from the point of view of an accelerated observer, is a particle state described by a density matrix at the temperature
\be T\,=\,\frac{a}{2\,\pi\,k_{B}} \label{eq:T} \ee called the Unruh-Davies temperature,where $a$ is the (constant) acceleration of the observer and $k_{B}$ is the Boltzmann constant; 2)an accelerated observer in the empty Minkowski space will detect a thermal bath of particles at the temperature ~\ref{eq:T}.

As was stressed in ~\cite{Bel1}~\cite{Bel2}~\cite{Bel3}~\cite{Bel4} (in
particular ~\cite{Bel4} is recommended for an especially clear
account) the crucial point here is not that an
accelerator detector (observer) would in someway react to the vacuum
state of the field in Minkowski space, rather that this response is
universal, i.e. independent from the structure of the detector itself,
from the quantum field considered, and from the details of the
interaction between them.
This situation could remind the response of a probe massive body to
the gravitational field, which is indeed universal. This universality
is in this case determined by the pure geometrical nature of the
gravitational field, so that the interaction between a body and the
gravity is determined uniquely by the geometry of the spacetime in
which the body moves, and not by its inner structure. If a similar
universality would appear in the different context of accelerated
observer in Minkowski space, this would mean that here only the
quantum properties of Minkowski vacuum state matter.
   
Consequently, two problems are involved here, which in principle are
different, but are claimed to be equivalent in the literature: the
physical properties of a quantum field when restricted to a
submanifold (the Rindler Space, (RS)) of Minkowski space (MS); the
behaviour of a constantly accelerated particle detector in empty flat
space.

 The first one deals only with basic principles of quantum theory and
 appears to be more fundamental, whether the second one should in
 general involve  also a description of structure and characteristics
 of the detector, and details of interaction with the quantum
 field. Again, it is the analysis of the first problem that made possible to claim for universality of the Unruh-like detector response.

 We will treat here only the first problem above, which is a
 particular example of the analysis of the behaviour of a field in a
 submanifold of a maximally analytically extended manifold.
% and also because it seems to be related in many ways with the phenomenon of q
%uantum evaporation of black holes, i.e. the Hawking effect
 %~\cite{Unruh}~\cite{Hawking}~\cite{DesLev}~\cite{CadoMign}~\cite{Wald}~\cite{%BirDav}.

  The procedure used by Unruh is based on a quantization scheme for a free field in MS, alternative but claimed to be equivalent to the standard one, which uses as the Hilbert space of solutions of the wave equation  
 \be \mathcal{H}_{U}\,=\,\mathcal{H}_{R}\,\oplus\,\mathcal{H}_{L} \ee
 where $\mathcal{H}_{R}$ consists of solutions which are non-zero
 everywhere but in the L sector, which have positive frequency with
 respect to the \lq\lq Rindler time" $\eta$, and which reduce
 themselves to the well known Fulling modes ~\cite{Fulling}, and
 $\mathcal{H}_{L}$ is given by the solutions which are non-zero
 everywhere but in R and with negative frequency with respect to
 $\eta$. Then it is obtained a representation of the Minkowski vacuum
 state as a state in $\mathcal{F}_{s}(\mathcal{H}_{U})$, i.e. the Fock
 space constructed on $\mathcal{H}_{U}$. Finally it is derived the
 particle content of Minkowski vacuum and the expression for the
 density matrix associated to it, when expressed as a mixed state in
 the Fock space $\mathcal{F}_{s}(\mathcal{H}_{R})$ having traced out
 the degrees of freedom related to the L sector, which is unaccessible
 to the Rindler observer.

The usual explanation of the Unruh effect is based exactly on the
presence, for a Rindler observer, which is confined inside the Rindler
wedge, of an event horizon which prevent him from having part of the
informations about the quantum field, so that he sees Minkowski vacuum
state as a mixed state. But this explanation is (as was indicated in
~\cite{Bel1}~\cite{Bel2}~\cite{Bel3}~\cite{Bel4} and also in our opinion) not
entirely satisfying, for several reasons: on the one hand, the
existence of horizons is due to overidealization of the problem, since
for physical accelerations (which last finite amount of time) no
horizon should be present, so that the response of an accelerated
detector, if of the Unruh-Davies type, cannot be caused by it; on the
other hand,
in the purely quantum theoretical treatment of the problem, we would
expect that the presence of event horizons affects deeply the fields,
from the point of view of Rindler observer, in the form of some kind
of boundary condition, which instead are totally absent in the Unruh
scheme, and in the usual quantization in RS. 

 What we are going to do in this paper is to extend to the spinor
 field the results
 obtained in ~\cite{Bel1}~\cite{Bel2}~\cite{Bel3}~\cite{Bel4} for the
 scalar field and to show the conditions to
 (and only to) which the quantum theory of the spinor field in the
 Rindler spacetime is well posed, to study the relationship between
 this construction and the usual one in MS, in order to find which
 role is played by these conditions in the derivation and
 interpretation of the Unruh effect in the spinor case, to analyse the
 Unruh quantization scheme and to understand finally what is its
 physical significance.

  What we will find is that a correct quantization procedure for the spinor field in Rinlder space requires the boundary condition
 \be \lim_{\rho\,\rightarrow\,0}\,\rho^{\frac{1}{2}}\,\Psi\left( 0\,,\,\rho\right)\,=\,0 \label{eq:cond} \ee
  i.e. the field should not grow up too rapidly at the origin of
 Rindler (and Minkowski) space. From a more general point of view, this means that the quantization on a background manifold which is not maximally extended requires a boundary condition which is absent in the quantization procedure over the extended manifold; in the literature this is not recognized clearly enough, and the usual procedure is to restrict the fields just considering a smaller domain of definition.

  Moreover, we show that also for the spinor field the Unruh quantization scheme is not valid in
  the whole Minkowski space, but only in a sector of it, namely the
  double wedge $R\cup L$. Consequently, the Unruh quantization implies the same kind of boundary condition, and cannot be used as a proof of the Unruh effect.

  These results, as we said, represent a generalization to the spinor field of
similar ones obtained for the scalar field (and also recently extended
also to  the electromagnetic
field ~\cite{Colosi}); consequently, they appear to be consequences of very general properties of Quantum Field Theory, and seem to be firmly established.
It is worth to note here that the Unruh effect is often identified
(mainly by mathematical physicists)
with the so-called Bisognano-Wichmann Theorem, in the context of the
algebraic approach to quantum field theory. We won't deal in this
paper with the algebraic approach, but the interested reader could
found the analog of our result for the scalar case extended to the
algebraic framework in ~\cite{Bel2} ~\cite{Bel3}. There it is shown
that the physical interpretation of the Bisognano-Wichmann Theorem, in
terms of accelerated observers in MS occurs in the same kind of
problems encountered in the conventional approach, that we will now discuss. 

  In addition to the main results cited, we obtained some minor but original results, necessary to achieve the first, namely: the explicit expression of the Lorentz boost generator (or Lorentz Momentum) for the spinor field, its eigenfunctions and their analytical representation holding in the whole MS (except for the origin), which is in turn a generalization of the Gerlach's Minkowski Bessel Modes ~\cite{Gerlach}.

  \section{Rindler Spacetime} \label{sec:Rind}
 Let's consider a particle moving in MS with constant acceleration $a$ along the x-axis; it will follow the trajectory given by (parameter $\tau$):
 \be t\,=\,a^{-1}\sinh(a\tau)\;\;\;\;\;x\,=\,a^{-1}\cosh(a\tau)\;\;\;y\,=\,y(0)\;\;\;z\,=\,z(0) \label{eq:acc} \ee
 This is an hyperbola in the (t,x) plane. The lines $t=\pm x$ represent asymptotes for it and event horizons for the moving particle. Varying $a$ we obtain different hyperbolas with the same characteristics. Let's now perform, starting from the Minkowski metric, the change in coordinates given by:
 \bea  t\,=\,\rho\,\sinh\,\eta\;\;\;\;\;\;x\,=\,\rho\,\cosh\,\eta \label{eq:cooRin} \label{eq:coo}\\
  \rho\,=\,\sqrt{x^{2}-t^{2}}\;\;\;\;\;\;\eta\,=\,arctgh\left(\frac{t}{x}\right) \eea          
 with the other coordinates left unchanged.
  The metric assumes the form:
 \be ds^{2}\,=\,\rho^{2}d\eta^{2}\,-\,d\rho^{2}\,-\,dy^{2}\,-\,dz^{2}  \label{Rindler} \ee
 Note that it describes a stationary spacetime. The worldlines $\rho=const, y=const, z=const$ correspond to uniformly accelerated observers, with $a=\rho^{-1}$ and proper time $\tau=\rho\eta$, as it can be seen comparing \ref{eq:acc} and \ref{eq:coo}. We can think at the Rindler Space as the collection of these worldlines, and this is the reason why RS is generally regarded as the \lq\lq natural\rq\rq manifold in which to describe accelerate motion. The hypersurfaces $\eta=const$ describe events which are simultaneous from the point of view of a \lq\lq Rindler (uniformly accelerated) observer". The Rindler manifold cannot be extended to negative values of $\rho$ trough $\rho=0$, for RS is no more rigid beyond the limit $\rho\rightarrow 0$. This hypersurface, in fact, represents, as we said, an event horizon for Rindler observers, which cannot see any event located beyond it. Nevertheless the horizon is a regular surface, of course, and the metric singularity is due only to the choice of the coordinates.

  The (~\ref{eq:cooRin}) cover only a sector of the whole MS and the others are covered by the following charts:
 \begin{center}
 {L\,\,:}\be t\,=\,\rho\,\sinh\,\eta\;\;\;x\,=\,\rho\,\cosh\,\eta \ee
 \be \rho\,=\,-\,\sqrt{x^{2}-t^{2}}\;\;\;\eta\,=\,arctgh(\frac{t}{x})
 \ee
{F\,\,:}
 \be t\,=\,\rho\,\cosh\,\eta\;\;\;x\,=\,\rho\,\sinh\,\eta \ee
 \be \rho\,=\,\sqrt{t^{2}-x^{2}}\;\;\;\eta\,=\,arctgh(\frac{x}{t}) \ee
 {P\,\,:}
 \be t\,=\,\rho\,\cosh\,\eta\;\;\;x\,=\,\rho\,\sinh\,\eta \ee
 \be \rho\,=\,-\,\sqrt{t^{2}-x^{2}}\;\;\;\eta\,=\,arctgh(\frac{x}{t}) \ee
 \end{center}
% It's important to note also that these characteristics of RS inside MS are an%alogous to that of Schwarzchild spacetime inside the Kruskal one, from both ge%ometrical and physical point of view.
          
 \section{Quantization in Rindler space} \label{sec:RQFT}
 We now turn to the problem of quantization of the spinor field in the Rindler wedge. The procedure will be the standard one, so we will first solve the Dirac equation looking for hamiltonian eigenfunctions.
 \subsection{The Dirac equation and its solutions in RS}
 Using the tetrad formalism, the Dirac equation in a generic curved
 spacetime is given by:
% ~\cite{BrilWhe}:
 \be (\,i\,\gamma^{\mu}\,\nabla_{\mu}\,-\,m\,)\,\Psi\,=\,0 \label{eq:D} \ee
 where: $\gamma^{\mu}\,=\,\theta^{\mu}_{\bar{\mu}}\,\gamma^{\bar{\mu}}$ are the analogous in curved spaces of the usual Dirac gamma matrices $\gamma^{\bar{\mu}}$, and they satisfy:
 \be
 \gamma^{\mu}\,\gamma^{\nu}\,+\,\gamma^{\nu}\,\gamma^{\mu}\,=\,g^{\mu\nu} \ee la  $\theta^{\mu}_{\bar{\mu}}$ is the inverse of the tetrad vector $\theta^{\bar{\mu}}_{\mu}$ with vectorial index $\mu$ and tetradic index $\bar{\mu}$; $\nabla_{\mu}=\partial_{\mu}-\Gamma_{\mu}$ is the spinorial covariant derivative, defined in such a way that $\nabla_{\mu}\Psi$ is a covariant vector;\be \Gamma_{\mu}\,=\,\frac{1}{8}\,\theta_{\bar{b}}^{b}\,\theta_{\bar{c}\,b\,;\,\mu}\,[\gamma^{\bar{c}},\gamma^{\bar{b}}] \ee is the spinorial connection.

  Let's consider the particular case of the Rindler metric:
 \be ds^{2}\,=\,\rho^{2}\,d\eta^{2}\,-\,d\rho^{2}\,-\,dy^{2}\,-\,dz^{2} \label{eq:metrica} \ee
 
%Consequently 
 The Dirac equation 
 %in Rindler space 
 assumes the form:
 \bea \left(
 i\,\partial_{\eta}+i\,\rho\,\gamma^{0}\gamma^{1}\partial_{\rho}+i\,\rho\,\gamma^{0}\gamma^{2}\partial_{y}+i\,\rho\,\gamma^{0}\gamma^{3}\partial_{z}+\frac{i}{2}\gamma^{0}\gamma^{1}-m\,\rho\,\gamma^{0} \right) \Psi=0 \\ \Longrightarrow \;\;\;\;\;\;\;\;i\,\partial_{\eta}\,\Psi\,=\left(\,-\,i\,\rho\,\alpha_{i}\,\partial_{i}\,-\,\frac{1}{2}\,i\,\alpha_{1}\,+\,m\,\rho\,\beta \right) \Psi \eea
that is a Shroedinger-like form with an hamiltonian given by:
\bea
 H_{R}\,=\,-\,i\,\rho\,\alpha_{i}\,\partial_{i}\,-\,\frac{1}{2}\,i\,\alpha_{1}\,+\,m\,\rho\,\beta \,= \;\;\;\;\;\;\;\;\;\;\;\;\;\;\;\;\;\;\;\;\;\;\;\;\;\;\;\;\;\;\;\;\\ =\,-\,i\,\rho\,\gamma^{0}\,\gamma^{1}\,\partial_{\rho}\,-\,i\,\rho\,\gamma^{0}\,\gamma^{2}\,\partial_{y}\,-\,i\,\rho\,\gamma^{0}\,\gamma^{3}\,\partial_{z}\,-\,\frac{1}{2}\,i\,\gamma^{0}\,\gamma^{1}\,+\,m\,\rho\gamma^{0} 
\label{eq:H} \eea
 We now look for solutions of the Dirac equation which are simultaneously eigenfunctions of the Rindler hamiltonian (~\ref{eq:H}) and of the operators $P_{y}$ and $P_{z}$ (of course they should also have the correct behaviour at infinity). So we expect to find a degeneracy of these solutions, because we know that three operators are not sufficient to completely characterize the states of the spinor field. This degeneracy is however of no relevance for our pourposes, so we will not deal with it.

  The solutions are:
 \be \Psi_{1\mathcal{M}}^{R}\,=\,N_{\mathcal{M}}\,\left( X_{1}^{R}\,K_{i\,\mathcal{M}-\frac{1}{2}}(\kappa\rho)\,+\,Y^{R}_{1}\,K_{i\,\mathcal{M}+\frac{1}{2}}(\kappa\rho) \right)\,e^{-\,i\,\mathcal{M}\,\eta}\,e^{i\,k_{2}\,y\,+\,i\,k_{3}\,z}\label{eq:Psi1} \ee

  with: 

  \be X^{R}_{1}\,=\,\left( \begin{array}{c} k_{3} \\
i\,\left(k_{2}\,+\,i\,m \right) \\ i\,\left(k_{2}\,+\,i\,m \right) \\
k_{3} \end{array} \right)
\;\;\;\;\;\;\;\;\;\;\;\;\;\;Y^{R}_{1}\,=\,\left( \begin{array}{c} 0 \\
i\,\kappa \\ -\,i\,\kappa \\ 0 \end{array} \right) \label{eq:XY1} \ee

and

\be \Psi^{R}_{2\mathcal{M}}\,=\,N_{\mathcal{M}}'\,\left( X^{R}_{2}\,K_{i\,\mathcal{M}-\frac{1}{2}}(\kappa\rho)\,+\,Y^{R}_{2}\,K_{i\,\mathcal{M}+\frac{1}{2}}(\kappa\rho) \right)\,e^{-\,i\,\mathcal{M}\,\eta}\,e^{i\,k_{2}\,y\,+\,i\,k_{3}\,z} \label{eq:Psi2} \ee

  with:

  \be X^{R}_{2}\,=\,\left( \begin{array}{c} 0 \\ i\,\kappa \\ i\,\kappa \\ 0 \end{array} \right) \;\;\;\;\;\;\;\;\;\;\;\;\;\; Y^{R}_{2}\,=\,\left( \begin{array}{c} k_{3} \\ i\,\left(k_{2}\,-\,i\,m \right) \\ -\,i\,\left(k_{2}\,-\,i\,m \right) \\ -\,k_{3} \end{array} \right) \label{eq:XY2} \ee

   where $k_{2}$,$k_{3}$ are eigenvalues of $P_{2}$, $P_{3}$, $\kappa=\sqrt{k_{2}\,+\,k_{3}\,+\,m^{2}}$, $\mathcal{M}$ is the eigenvalue of the hamiltonian, and $N_{\mathcal{M}}$, $N'_{\mathcal{M}}$ are normalization factors which are found to be:
 \be N_{\mathcal{M}}\,=\,N_{\mathcal{M}}'\,=\,\frac{1}{4\,\pi^{2}\,\sqrt{\kappa}}\,\sqrt{\cosh{\pi\,\mathcal{M}}} \ee
 \subsection{Physical characterization of the solutions}
 We now want to understand better the physical nature of the solutions (~\ref{eq:Psi1})(~\ref{eq:Psi2}), and this means to characterize them in a clearer way than just saying they are eigenfunctions of the Rindler hamiltonian.

  For this pourpose we need to find the expressions for the solutions
  in minkowski coordinates.

  In order to do it, it is necessary to consider carefully the way in
which spinors transform under coordinate transformations. 

It is well known that also in curved spacetimes spinors are
 characterized by their transformation properties under the action  of
 the Lorentz group, but restricting the attention to the local
 minkowskian neighbour of the point in which the spinors have to be
 calculated, i.e. considering local Lorentz transformations on the
 tangent space of each point. Consequently, under a general coordinate
 transformation, the spinor will undergo a Lorentz transformation, but
 with a \lq\lq velocity parameter\rq\rq which will be a function of
 the coordinates; this local Lorentz transformation has to be
 determined linearizing the coordinate transformation in which we are
 interested. In our case (transformation between Minkowskian and
 Rindler coordinate systems ~\ref{eq:cooRin}) the result is that the spinor transformation is given by:
 \be \Psi(t\,,\,x)\,=\,S(\eta)\,\Psi(\eta\,,\,\rho)\,=\,\exp{\left( \frac{1}{2}\,\gamma^{0}\,\gamma^{1}\,\eta\right)}\,\Psi(\eta\,,\,\rho) \ee

  Note that the operator we found, $S=\exp{\left( \frac{1}{2}\gamma^{0}\gamma^{1}\eta\right)}$, has the form of an operator resulting from a Lorentz coordinate transformation, with \lq\lq velocity" parameter $\eta$.

  The solutions (~\ref{eq:Psi1})(~\ref{eq:Psi2}) in minkowskian coordinates take the form: 
 \bea \Psi_{i,\mathcal{M}}^{R}(t,x,y,z)\,=\,N_{\mathcal{M}}\,\left(
 X^{R}_{i}\,K_{i\,\mathcal{M}-\frac{1}{2}}(\kappa\rho)\,e^{-\,\left(i\,\mathcal{M}-\frac{1}{2}\right)\eta}\,+\right. \nonumber \\ \left.+\,Y^{R}_{i}\,K_{i\,\mathcal{M}+\frac{1}{2}}(\kappa\rho)\,e^{-\,\left(i\,\mathcal{M}+\frac{1}{2}\right)\eta} \right)\,e^{i\,k_{2}\,y\,+\,i\,k_{3}\,z} \eea

  It must be clear that these are again defined only in the RS, but expressed in minkowskian coordinates, and this is why the normalization factor is still $N$.

  Now we can turn to the anticipated physical characterization of the
solutions: they are found to be eigenfunctions of the Boost Generator
Operator, or Lorentz Momentum. This could be aspected, because
1)$\Psi_{i\mathcal{M}}$ were eigenfunctions of the Rindler Hamiltonian
$H_{R}$ and
2)$H_{R}$ is precisely the Lorentz Boost Generator
written in Rindler coordinates, since
(3)the time evolution of a Rindler (uniformly accelerated) observer is properly a infinite succession of infinitesimal boost transformations.

  It is easy to verify this statement, once known the Lorentz Momentum operator, and this in turn can be obtained from the classical theory of fields.

  In fact, given the conserved quantity:
\be M^{0 i}\,=\,\int\,d^{3}x\,\left[ \left( x^{0}\,T^{i 0}\,-\,x^{i}\,T^{0 0} \right) \,+\,S^{0 i 0} \right] \ee
 corresponding to the invariance of the Lagrangian under boost transformations along the i-axis, (which are isometries of the Rindler spacetime), the explicit calculation of $T^{\mu\nu}$ and of $S^{\mu\nu}$ gives:
 \be M^{01}\,=\,\int\,d^{3}x\,\Psi^{\dagger}\,\left[ i\,t\,\partial^{i}\,+\,x^{i}\,\left( i\,\gamma^{0} \,\gamma^{i} \,\partial_{i} \,-\,m\,\gamma^{0} \right) +\, \frac{i}{2}\,\gamma^{0}\,\gamma^{i}\right]\,\Psi \label{eq:Mcl} \ee
 Interpreting it as a mean value of a quantum operator, we obtain for the Lorentz Momentum the expression (for contravariant and covariant components):
 \be  M^{0 i}\,=\,-\,i\,t\,\partial_{i}\,-\,x_{i}\,\left(
-\,i\,\gamma^{0}\,\gamma^{j}\,\partial_{j}\,+\,m\,\gamma^{0} \right)
+\,\frac{i}{2}\,\gamma^{0}\,\gamma^{i} \ee
 \be  M_{0
i}\,=\,+\,i\,t\,\partial_{i}\,+\,x_{i}\,\left(
-\,i\,\gamma^{0}\,\gamma^{j}\,\partial_{j}\,+\,m\,\gamma^{0} \right)
-\,\frac{i}{2}\,\gamma^{0}\,\gamma^{i} \label{eq:M} \ee
%where the symbols $\partial_{j}$ are intended to represent just partial deriva%tives, with no particular covariant characterization.

  Given this expression it can be verified that our solutions are eigenfunctions of the operator $M_{01}$ with eigenvalue $\mathcal{M}$, and this represent their physical characterization.
 \subsection{The second quantization in RS}
 We now possess all the necessary elements to perform the second quantization of the spinor field in RS, i.e. a set of normalized functions which are solutions of the equation of motion.
  We then expand the field in terms of them:
 \bea
 \lefteqn{\Psi^{R}(\eta,\rho,y,z)=\sum_{i=1,2}\int_{-\infty}^{+\infty}d\mathcal{M}\int_{-\infty}^{+\infty}dk_{2}\int_{-\infty}^{+\infty}dk_{3}\,a_{\mathcal{M}i}(k_{2},k_{3})\Psi_{i,\mathcal{M},k_{2},k_{3}}(t,x,y,z)=} \nonumber \\ &=& \sum_{i=1,2}\int_{0}^{+\infty}d\mathcal{M}\int_{-\infty}^{+\infty}dk_{2}\int_{-\infty}^{+\infty}dk_{3}\times \nonumber \\ & \times& \left( a_{\mathcal{M},i}(k_{2},k_{3})\Psi^{R}_{i,\mathcal{M},k_{2},k_{3}}(\eta,\rho,y,z)+b^{\dagger}_{\mathcal{M},i}(k_{2},k_{3})\Psi^{R}_{i,-\mathcal{M},k_{2},k_{3}}(\eta,\rho,y,z)\right)\;\;\;\;\;\;\;  \label{eq:exp} \eea
The second quantization is now performed considering the coefficients $a_{i,\mathcal{M}}$ and $b_{i,\mathcal{M}}$ (and their hermitian coniugated) as operators, and requiring for their anticommutators:
 \bea \{ a_{\mathcal{M},i}(k_{2},k_{3})\,,\,a_{\mathcal{M}',j}^{\dagger}(k'_{2},k'_{3})\}\,=\,\delta_{ij}\,\delta(\mathcal{M}-\mathcal{M}')\,\delta(k_{2}-k'_{2})\,\delta(k_{3}-k'_{3})\;\;\; \\ \{ b_{\mathcal{M},i}(k_{2},k_{3})\,,\,b_{\mathcal{M}',j}^{\dagger}(k'_{2},k'_{3})\}\,=\,\delta_{ij}\,\delta(\mathcal{M}-\mathcal{M}')\,\delta(k_{2}-k'_{2})\,\delta(k_{3}-k'_{3})\;\;\;  \\ 
\{ a_{\mathcal{M},i}(k_{2},k_{3})\,,\,b_{\mathcal{M}',j}^{\dagger}(k'_{2},k'_{3})\}\,=\,0\;\;\;\;\forall\;i,j,\mathcal{M},\mathcal{M}',k_{2},k_{2}',k_{3},k_{3}'\;\;\;\;\;\;\;\;\;\;\;\; \\ \{ a_{\mathcal{M},i}(k_{2},k_{3})\,,\,b_{\mathcal{M}',j}(k'_{2},k'_{3})\}\,=\,0\;\;\;\;\forall\;i,j,\mathcal{M},\mathcal{M}',k_{2},k_{2}',k_{3},k_{3}'\;\;\;\;\;\;\;\;\;\;\;\; \eea 

  and the quantum states of the field are constructed from the Rindler vacuum state $\mid 0\,\rangle_{R}$, defined by:
 \be a_{\mathcal{M},i}(k_{2},k_{3})\,\mid 0\,\rangle_{R}\,=\,0\;\;\;\; \forall\;\;i,\,\mathcal{M},\,k_{2},\,k_{3} \ee

  Now we will study if this quantum construction is well posed and at
which conditions.

\subsection{Conditions for the quantization in RS}
Let's consider again the Rindler hamiltonian:
 \bea \lefteqn{H_{R}\,=\,-\,i\,\rho\,\alpha_{i}\,\partial_{i}\,-\,\frac{1}{2}\,i\,\alpha_{1}\,+\,m\,\rho\,\beta \,=} \\ &=&\,-\,i\,\rho\,\gamma^{0}\,\gamma^{1}\,\partial_{\rho}\,-\,i\,\rho\,\gamma^{0}\,\gamma^{2}\,\partial_{y}\,-\,i\,\rho\,\gamma^{0}\,\gamma^{3}\,\partial_{z}\,-\,\frac{1}{2}\,i\,\gamma^{0}\,\gamma^{1}\,+\,m\,\rho\gamma^{0}\;\;\;\;\;\;\;\;\;\;\;  \eea
 and let's check whether it represents an hermitian operator, that is a necessary and sufficient condition for the completeness and orthonormality of the modes used. We should verify the condition
 \be \left( H_{R}\Phi\,,\,\Psi\right)\,=\,\left(
 \Phi\,,\,H_{R}\Psi\right) \ee
with scalar product given by:
 \be \left(
 \Phi\,,\,\Psi\right)\,=\,\int d\Sigma_{\mu}\,\bar{\phi}\gamma^{\mu}\psi\,=\,\int_{-\infty}^{+\infty}\,dy\,\int_{-\infty}^{+\infty}\,dz\,\int_{0}^{+\infty}d\rho\,\Phi^{\dagger}\,\Psi \ee
The explicit calculation shows that:
\be \left( H_{R}\Phi\,,\,\Psi\right)\,=\,\left(
\Phi\,,\,H_{R}\,\Psi\right)\,+\,\int_{-\infty}^{+\infty}\,dy\,\int_{-\infty}^{+\infty}\,dz\,\left[ i\,\Phi^{\dagger}\,\alpha_{1}\,\rho\,\Psi \right]^{\rho\,=\,+\infty}_{\rho\,=\,0} \ee 
 Then it is evident that the hermiticity of the hamiltonian is assured if and only if 
  \be
 \lim_{\rho\rightarrow\,0}\,\rho^{\frac{1}{2}}\,\Psi\left(\eta\,,\,\rho\,,\,y\,,\,z\right)\,=\,0 \;\;\;\;\forall\eta \label{eq:Condizione} \ee
and of course with analogous condition at $\rho\rightarrow +\infty$,
i.e. the usual requirement of vanishing of the fields at spatial infinity.

  We emphasize that, since the field $\Psi$ is to be considered as an operator-valued distribution, this condition should be interpreted in the weak sense, this meaning that every matrix element of the quantity $\rho^{\frac{1}{2}}\Psi$ calculated with respect to any pair of physical states has to go to zero as $\rho\rightarrow 0$.

  We stress again that a similar condition was already found for the
  scalar field in RS ~\cite{Bel1}~\cite{Bel2}~\cite{Bel3}, and for the
  vector field ~\cite{Colosi}. 

  If the condition (~\ref{eq:Condizione}) is necessary for the
hermiticity of the hamiltonian, it is expected to appear also in the
analysis of the coefficients of the expansion (~\ref{eq:exp}). So we
write the explicit expression of the coefficients
$a_{i,\mathcal{M}}(k_{2},k_{3})$:
\bea \lefteqn{a_{\mathcal{M},i}(k_{2},k_{3})\,=\,\left( \Psi_{i,\mathcal{M},k_{2},k_{3}}\,,\,\Psi \right)_{R}\,=} \\ &=& +\,\int_{-\infty}^{+\infty}\,dy\,\int_{-\infty}^{+\infty}\,dz\,\int_{0}^{+\infty}d\rho\,N_{\mathcal{M}}^{*}\left[ \left( X_{i}^{\dagger}\,K_{i\mathcal{M}+\frac{1}{2}}\,+\,Y_{i}^{\dagger}\,K_{i\mathcal{M}-\frac{1}{2}}\right)\,\Psi\right]\,\times \nonumber \\ &\times& e^{(i\,\mathcal{M}\,\eta\,+\,i\,k_{2}\,y\,+\,i\,k_{3}\,z)} \label{eq:coeff} \eea

and consider the behaviour of the $K_{i\mathcal{M}\pm\frac{1}{2}}(\kappa\rho)$ for $\rho\simeq 0$. 
 
 We have:
 \bea \lefteqn{K_{i\mathcal{M}\pm\frac{1}{2}}(\kappa\rho)\,\simeq\,\frac{\pi}{2}\frac{1}{\sin\left(\pi\left( i\mathcal{M}\pm\frac{1}{2}\right)\right)}\times} \nonumber \\ &\times&\left[ \frac{\kappa^{-\left( i\mathcal{M}\pm\frac{1}{2}\right)}}{\Gamma\left( -\left( i\mathcal{M}\pm\frac{1}{2}\right)\,+\,1\right)}\left( \frac{\rho}{2}\right)^{-\left( i\mathcal{M}\pm\frac{1}{2}\right)}\,-\,\frac{\kappa^{\left( i\mathcal{M}\pm\frac{1}{2}\right)}}{\Gamma\left( \left( i\mathcal{M}\pm\frac{1}{2}\right)\,+\,1\right)}\left( \frac{\rho}{2}\right)^{\left( i\mathcal{M}\pm\frac{1}{2}\right)}\right]\;\;\;\;\;\;\;\,\;\;\;\;\;\; \label{eq:K0}\eea
 We see that a divergence is present for $\rho\rightarrow 0$ (we note also that, for $\mathcal{M}=0$, we have the exact expression:
 \be
K_{\frac{1}{2}}(\kappa\rho)\,=\,\sqrt{\frac{\pi}{2\,\kappa\,\rho}}\,e^{-\,\kappa\,\rho}
\ee
again with the same type of divergence).

  It is so evident that, in order the integral (~\ref{eq:coeff}) to
  converge, is necessary to require the boundary condition
  (~\ref{eq:Condizione}) on the field. Otherwise, the coefficients
  $a_{i,\mathcal{M}}(k_{2},k_{3})$ are not defined and consequently
  the fundamental operators of quantum field theory like energy or
  particle number, which are built in terms of the annihilation and
  creation operators, are similarly not defined.

We want now to prove the necessity of the condition
(~\ref{eq:Condizione}) in an even more apparent way, i.e. showing that
the requirement of finiteness of the mean value of the
energy in a generic state implies indeed (~\ref{eq:Condizione}).
We work for simplicity in the plane $(\eta,\rho)$
Consider the state of the field $\mid g \rangle$ given by 
\be \mid g \rangle\,=\,c^{\dagger}(g)\,\mid
0\rangle_{R}\,=\,\sum_{i}\int_{0}^{\infty}\frac{d\mathcal{M}}{\mathcal{M}^{\frac{1}{2}}}\,g(\mathcal{M})\,c_{i,\mathcal{M}}^{\dagger}\,\mid
0\,\rangle_{R}\label{eq:g} \ee
i.e. a generic sovrapposition of eigenstates of the hamiltonian with
weight function $g(\mathcal{M})$, and consider also the related
one-particle amplitude for the field in RS given by
\be \Psi^{R}_{g}\,=\,_{R}\langle 0 \mid\, \Psi_{R}\,\mid
g\,\rangle\,=\,e^{-i\,H_{R}\,\eta}\,\phi_{g} \ee

where 
\be
\phi_{g}\,=\,\sum_{i}\int_{0}^{\infty}\frac{d\mathcal{M}}{\mathcal{M}^{\frac{1}{2}}}\frac{g(\mathcal{M})}{4\pi^{2}\sqrt{\kappa}}\sqrt{\cosh{\pi\mathcal{M}}}\left(X_{i}^{R}\,K_{i\mathcal{M}-\frac{1}{2}}\,+\,Y_{i}^{R}\,K_{i\mathcal{M}+\frac{1}{2}}\right)\label{eq:phig}
\ee

Consequently we have the following translation of physical requirements
(normalization of the states and finiteness of the mean value of the
energy) into
mathematical requirements on the weight function $g(\mathcal{M})$:
\be \langle g \mid g
\rangle\,=\,\int_{0}^{\infty}\frac{d\mathcal{M}}{\mathcal{M}}\mid
g(\mathcal{M})\mid ^{2}\,=\,1 \label{eq:nor}\ee 

\bea \lefteqn{ \langle g \mid H \mid
g\rangle\,=} \\ &=&\,\int_{0}^{\infty}d\rho\,\phi_{g}^{\dagger}H\phi_{g}\,=\,\int_{0}^{\infty}d\rho\,\phi_{g}^{\dagger}\left\{
-i\rho\alpha_{1}\partial_{\rho}-\frac{i}{2}\alpha_{1}+m\rho\beta\right\}\phi_{g}\,=\label{eq:valmedH}\\
&=&\,2\,\int_{0}^{\infty}d\mathcal{M}\mid
g(\mathcal{M})\mid^{2}\,<\,\infty \label{eq:meanvalue}\eea

From the equation (~\ref{eq:valmedH}) it is not immediately manifest
what should be the behaviour of the funtions $\phi_{g}$ with respect
to $\rho$, and in order to understand it we have to analyse directly
the expression (~\ref{eq:phig}), but taking into proper account the
physical requirements (~\ref{eq:nor}) and (~\ref{eq:meanvalue}).

Let`s consider the quantity $\rho^{\frac{1}{2}}\phi_{g}$ for
$\rho\rightarrow 0$, using the expression ~\ref{eq:phig} and the
formula ~\ref{eq:K0} for the modified Bessel function for
$\rho\rightarrow 0$. Simple manipulations lead to
 
\bea \lefteqn{ \rho^{\frac{1}{2}}\,\phi_{g}\,\simeq_{\rho\rightarrow
0}\,\sum_{i}\,\int_{0}^{\infty}\frac{d\mathcal{M}}{\mathcal{M}^{\frac{1}{2}}}\,N_{\mathcal{M}}\,g(\mathcal{M})\,\frac{\pi}{\sqrt{2\kappa}}\times}
\\ &\times&
\left\{ -X_{i}^{R}\frac{1}{ \sin( \pi ( i\mathcal{M}-\frac{1}{2}))
\Gamma\left(\frac{1}{2}+i\mathcal{M}\right)}
\left(\frac{\kappa\rho}{2}\right)^{i\mathcal{M}}\,+\right. \nonumber \\
&+& \left. Y_{i}^{R}\frac{1}{\sin\left(\pi\left(i\mathcal{M}+\frac{1}{2}\right)\right)\Gamma\left(\frac{1}{2}-i\mathcal{M}\right)}\left(\frac{\kappa\rho}{2}\right)^{-i\mathcal{M}}\right\}\,=
\nonumber
\\
&=& \sum_{i}\left\{
-\,X_{i}^{R}\,G_{A}\left(\rho,\kappa\right)\,+\,Y_{i}^{R}\,G_{B}\left(\rho,\kappa\right)\right\}
\eea

Now we want to study in details these quantities $G_{A}$ and $G_{B}$;
  we will do an explicit calculation only for the first one, since the
  argument for the second one is analogous.

First let's write $G_{A}$ as:
\bea G_{A}\,=\,\int_{0}^{\infty}
\frac{d\mathcal{M}}{\mathcal{M}^{\frac{1}{2}}}\,
N_{\mathcal{M}}\,g(\mathcal{M})
\,\frac{\pi}{\sqrt{2\kappa}}
\frac{1}{ \sin( \pi ( i\mathcal{M}-\frac{1}{2}))
\Gamma\left(\frac{1}{2}+i\mathcal{M}\right)}
\left(\frac{\kappa\rho}{2}\right)^{i\mathcal{M}}\,= \nonumber \\
 =\, G_{A1}\,+\,G_{A2}\,+\,G_{A3} \eea

where we have just splitted the integration domain into three parts,
i.e.
$(0,\infty)=(0,\mathcal{M}_{1})\cup(\mathcal{M}_{1},\mathcal{M}_{2})\cup(\mathcal{M}_{2},\infty)$,
with $\mathcal{M}_{1}<< 1$ and $\mathcal{M}_{2} >> 1$.

Consider the term $G_{A3}$. Using the explicit form of the
normalization factor $N_{\mathcal{M}}$, the asymptotic expression for
the Gamma function (for $\mathcal{M}\rightarrow\infty$), and basic
formulas for hyperbolic functions, we obtain:

\be \mid G_{A3}
\mid^{2}\,\leq_{\mathcal{M}_{2}\rightarrow\infty}\,\frac{1}{32\pi^{3}\kappa^{2}\mathcal{M}_{2}^{3}}\int_{\mathcal{M}_{2}}^{\infty}d\mathcal{M}\mid
g(\mathcal{M})\mid^{2} \ee

now, given the finiteness condition for the mean value of the energy (~\ref{eq:meanvalue}), we have that
$G_{A3}$ could be made as small as we want with sensible choices of
$\mathcal{M}_{2}\rightarrow\infty$.

Consider the term $G_{A2}$. We can easily obtain the inequality:

\be \mid G_{A2}\mid \leq
\int_{\mathcal{M}_{1}}^{\mathcal{M}_{2}}d\mathcal{M}\,C(\mathcal{M})\,\mid
g(\mathcal{M}) \mid \ee 

where $C(\mathcal{M})$ is a non singular function in the
interval of integration. Taking into account the inequality $\mid
g(\mathcal{M})\mid \leq \frac{1}{2} ( 1 + \mid
g(\mathcal{M})\mid^{2})$, and the nomalization condition
(~\ref{eq:nor}), it is easy to see that the integral above should
converge.
Now applying the Riemann-Lebesgue Lemma, we conclude that $G_{A2}$
vanishes for $\rho\rightarrow 0$.

Coming to the term $G_{A1}$, we have:

\bea
G_{A1}\,=\,\int_{0}^{\mathcal{M}_{1}}
d\mathcal{M}\frac{1}{\mathcal{M}^{\frac{1}{2}}}\frac{1}{2^{\frac{5}{2}}\pi\kappa}\sqrt{\cosh\pi\mathcal{M}}\,
g(\mathcal{M})\frac{1}{\sin(\pi(i\mathcal{M}-\frac{1}{2}))\Gamma(i\mathcal{M}+\frac{1}{2})}\times
\nonumber \\ \times\left(\frac{\kappa\rho}{2}\right)^{i\mathcal{M}}\, \lesssim\,
-\,\frac{1}{2^{\frac{3}{2}}\pi^{\frac{3}{2}}\kappa}\int_{0}^{\infty}\frac{d\mathcal{M}}{\mathcal{M}^{\frac{1}{2}}}\,g(\mathcal{M})\left(\frac{\kappa\rho}{2}\right)^{i\mathcal{M}}
\eea

being $\mathcal{M}_{1}<<1$.
Let's restrict the calculation to the case in which $g(\mathcal{M})$
vanish with $\mathcal{M}\rightarrow 0$ as a suitable high power of $\mathcal{M}$,
i.e.
\be g(\mathcal{M})\simeq
a\,\mathcal{M}^{\alpha}\;\;\;\;\;\;\alpha\,\geq\,\frac{1}{2}\;\;\;\;\;\;\mathcal{M}\rightarrow
0 \ee
(the results could be generalised to other cases). Note that the
vanishing of the weight function is required also by the condition
(~\ref{eq:nor}).

We obtain the inequality:
\be
G_{A1}\,\leq\,\frac{a}{2^{\frac{3}{2}}\pi^{\frac{3}{2}}\kappa}\frac{1}{\ln\frac{\kappa\rho}{2}}  \ee
so that $ G_{A1}\,\rightarrow 0$ as $\rho\rightarrow 0$.

Consequently we have proved that for the generic physical state
(~\ref{eq:g}) to be normalised and to have finite energy, the boundary
condition (~\ref{eq:Condizione}) is necessary. In other words, since  (~\ref{eq:nor}) and (~\ref{eq:meanvalue}) imply $\rho^{\frac{1}{2}}\phi_{g}\rightarrow 0$ for $\rho\rightarrow 0$, we know that, if this condition is not satisfied, then (~\ref{eq:nor}) or (~\ref{eq:meanvalue}) doesn't hold, and consequently the state $\mid g\rangle$ is not a physical state.

  \subsection{Discussion}
 We have found that the quantum theory of the
 spinor field in Rindler space is well defined if and only if we have
 the condition (~\ref{eq:Condizione}). This condition means that the
 field has to be quantized in a different way in MS and RS, because
 the horizons are not only of a causal but also of a physical
 significance for it. It also means that the usual procedure of
 quantize the field in RS, which just restrict its domain of
 definition, is not correct, because this kind of restriction is not
 enough to have a well posed theory. If one would like to study the
 spinor field in RS and work with physical states and modes, the two
 possible ways are to construct suitable wave packets made as
 combinations of the modes $\Psi_{i\mathcal{M}}$, or to consider from
 the beginning a constrained hamiltonian, different from $H_{R}$ which
 automatically assures that the condition (~\ref{eq:Condizione}) is
 satisfied. The crucial point is however that this condition prevents
 any relationship between the quantization in RS and that in MS, and
 it means that RS should be treated as an manifold on its own, and not
 as a submanifold of MS (the Rindler wedge), i.e. the two quantizations define two different physical systems. This is because in
 the latter case  we would not be free to choose any boundary condition for the field at the origin, because the state of the field would be
 determined from the beginning as a state in $\mathcal{H}_{M}$ (the
 Minkowski vacuum state $\mid 0 \rangle_{M}$). Consequently, it is not
 possible to describe any state of the spinor field quantized in
 Minkowski space, and then restricted to the Rindler wedge, as a state
 in $\mathcal{H}_{R}$, and we could exspect that the necessity of boundary conditions on the field would manifest itself also in the analysis of the Unruh effect, which concerns exactly this relationship between RS and MS quantum constructions. To show this will be our next problem.  

 \section{Analysis of the Unruh effect}
 In order to analyse this relationship, it is convenient to perform a
 quantization in Minkowski spacetime which is different from the
 standard one (that in plane waves) and which could be more easily
 compared with the Rindler one; namely, a quantization in terms of the
 Lorentz Momentum eigenfunctions defined in the whole Minkowski space,
 that represents analytical continuation of the $\Psi_{i,\mathcal{M}}$
 we found before, spinorial anologous of the Fulling modes for the scalar
 field, and that reduce themselves to these when restricted to the
 Rindler wedge., A physical motivation for this choice could be seen in
 the fact that the trajectories of a Rindler observer are the orbits
 of the Lorentz group, and that the R sector of MS is left invariant
 by the action of this group.
 The determination of these functions will require some preliminary steps.   

\subsection{Solutions in the other sectors and their relationship}
 First of all we will study the relations between the solutions of Dirac equation in the different sectors F, L, P (see ~\ref{sec:Rind}).
 Of course, we omit the passages and just write down the form of Dirac equation in the sector and its possible solutions . That is the following.
 \begin{center}
 {\large F sector:}
 \end{center}
 \be  \left( i\,\gamma^{0}\,\partial_{\rho}\,+\,\frac{i}{\rho}\,\gamma^{1}\,\partial_{\eta}\,+\,\frac{i}{2}\,\frac{1}{\rho}\,\gamma^{0}\,+\,i\,\gamma^{2}\,\partial_{y}\,+\,i\,\gamma^{3}\,\partial_{z}\,-\,m\right)\,\Psi\,=\,0 \ee

  \be \Psi_{i\mathcal{M}}^{F,\,\pm}\,=\,M^{\pm}_{i}\left( X_{i}^{F}\,K_{i\mathcal{M}-\frac{1}{2}}(\pm\,i\,\kappa\,\rho)\,+\,Y_{i}^{F}\,K_{i\mathcal{M}+\frac{1}{2}}(\pm\,i\,\kappa\,\rho)\right)\,e^{-\,i\,\mathcal{M}\,\eta}\,e^{i\,k_{2}\,y\,+\,i\,k_{3}\,z} \ee

  with $i=1,2$ and
\be X_{1}^{F}\,=\,\left( \begin{array}{c} k_{3} \\
i\,\left(k_{2}\,+\,i\,m \right) \\ i\,\left(k_{2}\,+\,i\,m \right) \\
k_{3} \end{array} \right)
\;\;\;\;\;\;\;\;\;\;\;\;\;\;Y_{1}^{F}\,=\,\left( \begin{array}{c} 0 \\
\mp\,\kappa \\ \pm\,\kappa \\ 0 \end{array} \right) \ee
\be X_{2}^{F}\,=\,\left( \begin{array}{c} 0 \\ \pm\,\kappa \\ \mp\,\kappa \\ 0 \end{array} \right) \;\;\;\;\;\;\;\;\;\;\;\;\;\;Y_{2}^{F}\,=\,\left( \begin{array}{c} k_{3} \\ i\,\left(k_{2}\,-\,i\,m \right) \\ -\,i\,\left(k_{2}\,-\,i\,m \right) \\ -\,k_{3} \end{array} \right) \ee

  \begin{center}
 {\large L sector:}
 \end{center}
 \be \left(  i\,\frac{1}{\rho}\,\gamma^{0}\,\partial_{\eta}\,+\,i\,\gamma^{1}\,\partial_{\rho}\,-\,i\,\gamma^{2}\,\partial_{y}\,-\,i\,\gamma^{3}\,\partial_{z}\,+\,\frac{1}{2}\,i\,\frac{1}{\rho}\,\gamma^{1}\,+\,m \right) \Psi\,=\,0 \ee
 \be \Psi_{i\mathcal{M}}^{L,\,\pm}\,=\,O_{i}^{\pm}\,\left( X_{i}^{L}\,K_{i\,\mathcal{M}-\frac{1}{2}}(\kappa\rho)\,+\,Y_{i}^{L}\,K_{i\,\mathcal{M}+\frac{1}{2}}(\kappa\rho) \right)\,e^{-\,i\,\mathcal{M}\,\eta}\,e^{i\,k_{2}\,y\,+\,i\,k_{3}\,z} \ee
 with
 \be X_{1}^{L}\,=\,\left( \begin{array}{c} k_{3} \\ i\,\left(k_{2}\,+\,i\,m \right) \\ i\,\left(k_{2}\,+\,i\,m \right) \\ k_{3} \end{array} \right) \;\;\;\;\;\;\;\;\;\;\;\;\;\;Y_{1}^{L}\,=\,\left( \begin{array}{c} 0 \\ -\,i\,\kappa \\ i\,\kappa \\ 0 \end{array} \right) \ee
 \be X_{2}^{L}\,=\,\left( \begin{array}{c} 0 \\ -\,i\,\kappa \\
 -\,i\,\kappa \\ 0 \end{array} \right) \;\;\;\;\;\;\;\;\;\;\;\;\;\;
 Y_{2}^{L}\,=\,\left( \begin{array}{c} k_{3} \\
 i\,\left(k_{2}\,-\,i\,m \right) \\ -\,i\,\left(k_{2}\,-\,i\,m \right)
 \\ -\,k_{3} \end{array} \right)  \ee

\begin{center}
 {\large P sector:}
 \end{center}
 \be  \left( i\,\gamma^{0}\,\partial_{\rho}\,+\,\frac{i}{\rho}\,\gamma^{1}\,\partial_{\eta}\,+\,\frac{i}{2}\,\frac{1}{\rho}\,\gamma^{0}\,-\,i\,\gamma^{2}\,\partial_{y}\,-\,i\,\gamma^{3}\,\partial_{z}\,+\,m\right)\,\Psi\,=\,0 \ee

  \be \Psi_{i\mathcal{M}}^{P,\,\pm}\,=\,P_{i}^{\pm}\left( X_{i}^{P}\,K_{i\mathcal{M}-\frac{1}{2}}(\pm\,i\,\kappa\,\rho)\,+\,Y_{i}^{P}\,K_{i\mathcal{M}+\frac{1}{2}}(\pm\,i\,\kappa\,\rho)\right)\,e^{-\,i\,\mathcal{M}\,\eta}\,e^{i\,k_{2}\,y\,+\,i\,k_{3}\,z} \ee

  with
 \be X_{1}^{P}\,=\,\left( \begin{array}{c} k_{3} \\ i\,\left(k_{2}\,+\,i\,m \right) \\ i\,\left(k_{2}\,+\,i\,m \right) \\ k_{3} \end{array} \right) \;\;\;\;\;\;\;\;\;\;\;\;\;\;Y_{1}^{P}\,=\,\left( \begin{array}{c} 0 \\ \pm\,\kappa \\ \mp\,\kappa \\ 0 \end{array} \right) \ee
 \be X_{2}^{P}\,=\,\left( \begin{array}{c} 0 \\ \mp\,\kappa \\ \mp\,\kappa \\ 0 \end{array} \right) \;\;\;\;\;\;\;\;\;\;\;\;\;\;Y_{2}^{P}\,=\,\left( \begin{array}{c} k_{3} \\ i\,\left(k_{2}\,-\,i\,m \right) \\ -\,i\,\left(k_{2}\,-\,i\,m \right) \\ -\,k_{3} \end{array} \right) \ee

  As regards to the relationship between these solutions, this could be found by analytically continuing the functions across the event horizons, which represent branch points for these functions. Using the variables
 \be x_{+}\,=\,x\,+\,t \;\;\;\;\;\,\;\;\,\;\;x_{-}\,=\,t\,-\,x \ee
 the passages trough the horizons is given by the substitutions: $-x_{-}\rightarrow x_{-}e^{\pm i\pi}$ ($R\rightarrow F$) $x_{+}\rightarrow -x_{+}e^{\pm i\pi}$ ($F\rightarrow L$), $x_{-}\rightarrow-x_{-}e^{\pm i\pi}$ ($L\rightarrow P$), $-x_{+}\rightarrow x_{+}e^{\pm i\pi}$ ($P\rightarrow R$).
 The result is that the solutions of the Dirac equation are linked by
 two possible paths of analytical continuation, namely that one corresponding (we name it A) to the transformation $-x_{\pm}\rightarrow x_{\pm}e^{+ i\pi}$ 
 and linking in succession the functions
 \be \Psi_{i,\mathcal{M}}^{R}\,\rightarrow\,\Psi_{i,\mathcal{M}}^{F,\,+}\,\rightarrow\,\Psi_{i,\mathcal{M}}^{L,\,+}\,\rightarrow\,\Psi_{i,\mathcal{M}}^{P,\,-}\,\rightarrow\,\Psi_{i,\mathcal{M}}^{R} \ee
and that one corresponding to the transformation $-x_{\pm}\rightarrow x_{\pm}e^{- i\pi}$ (we name it B)
 and linking in succession the functions
 \be \Psi_{i,\mathcal{M}}^{R}\,\rightarrow\,\Psi_{i,\mathcal{M}}^{F,\,-}\,\rightarrow\,\Psi_{i,\mathcal{M}}^{L,\,-}\,\rightarrow\,\Psi_{i,\mathcal{M}}^{P,\,+}\,\rightarrow\,\Psi_{i,\mathcal{M}}^{R} \ee
 
 Moreover, it is possible to demonstrate that the normalization factors of the different solutions are related each one other in such a way that, once determined $N$, all the others are determined consequently.
 \subsection{Lorentz Momentum eigenfunctions in MS}
 We now turn to the problem of finding a unified representation for the eigenfunctions of the Boost Generator.
 Let's consider the integral representations of the Bessel functions $K_{\nu}(\rho)$ given by:
 \bea
 K_{\nu}(\rho)\,=\,\frac{1}{2}\,e^{-\,\frac{i\pi\nu}{2}}\,\int_{-\infty}^{+\infty}e^{i\,\rho\,\sinh\vartheta}\,e^{\nu\,\vartheta}\,d\vartheta \label{eq:rappr1}\\ K_{\nu}(\rho)\,=\,\frac{1}{2}\,e^{\frac{i\pi\nu}{2}}\,\int_{-\infty}^{+\infty}e^{i\,\rho\,\sinh\vartheta}\,e^{-\,\nu\,\vartheta}\,d\vartheta \eea
Using the second one to express in an integral form and in minkowski coordinates the functions $K_{\nu}(\rho(t,x))e^{-\nu\eta(t,x)}$ (using the coordinate transformation ~\ref{eq:cooRin}), we obtain:
\bea \lefteqn{K_{\nu}(\kappa\rho)\,e^{-\,\nu\,\eta}\,=} \\ &=&
\,\frac{1}{2}\,e^{\frac{i\pi\nu}{2}}\,\int_{-\infty}^{+\infty}e^{i\,\kappa\,\rho\,\sinh\vartheta}\,e^{-\,\nu\,\left(\vartheta\,+\,\eta\right)}\,d\vartheta\,=
 \\
 &=&  \frac{1}{2}\,e^{\frac{i\pi\nu}{2}}\,\int_{-\infty}^{+\infty}e^{i\,\kappa\,\rho\,\sinh\left(\vartheta\,-\,\eta\right)}\,e^{-\,\nu\,\vartheta}\,d\vartheta\,= \\ &=& \frac{1}{2}\,e^{\frac{i\pi\nu}{2}}\,\int_{-\infty}^{+\infty}e^{\left[i\,\kappa\,\rho\,\sinh\vartheta\,\cosh\eta\,-\,i\,\kappa\,\rho\,\cosh\vartheta\,\sinh\eta\right]}\,e^{-\,\nu\,\vartheta}\,d\vartheta\,= \\
  &=&
 \frac{1}{2}\,e^{\frac{i\pi\nu}{2}}\,\int_{-\infty}^{+\infty}e^{i\,\kappa\left[ x\,\sinh\vartheta\,-\,t\,\cosh\vartheta\right]}\,e^{-\,\nu\,\vartheta}\,d\vartheta\,= \label{eq:rap1}\\ &=& \frac{1}{2}\,e^{\frac{i\pi\nu}{2}}\,\int_{-\infty}^{+\infty}\,P_{\vartheta}^{-}(t\,,\,x)\,e^{-\,\nu\,\vartheta}\,d\vartheta \eea
where $P^{-}_{\vartheta}(t,x)$ represent (2-dim) positive frequency plane waves with $\omega=\kappa\cosh\vartheta$ and $k_{x}=\kappa\sinh\vartheta$.
 Using the first one, and with the same procedure, we have:
 \bea K_{\nu}(\kappa\rho)\,e^{-\,\nu\,\eta}\,=\,\frac{1}{2}\,e^{-\,\frac{i\pi\nu}{2}}\,\int_{-\infty}^{+\infty}e^{i\,\kappa\left[ x\,\sinh\vartheta\,+\,t\,\cosh\vartheta\right]}\,e^{\nu\,\vartheta}\,d\vartheta\,= \label{eq:rap2} \\ = \,\frac{1}{2}\,e^{-\,\frac{i\pi\nu}{2}}\,\int_{-\infty}^{+\infty}\,P_{\vartheta}^{+}(t\,,\,x)\,e^{\nu\,\vartheta}\,d\vartheta \eea
 where now $P_{\vartheta}^{+}(t\,,\,x)$ are (2-dim) plane waves with negative frequency $-\omega=-\kappa\cosh\vartheta$. What we found means that the functions $K_{\nu}(\rho(t,x))e^{-\nu\eta(t,x)}$ can be expressed equivalently as linear (integral) combination of positive or negative two dimensional plane waves.

  Inserting these formulas into the expression for $\Psi_{i,\mathcal{M}}^{R}(t,x,y,z)$, not taking care of the normalization factor, gives 
 \bea \Psi_{i\mathcal{M}}^{\mp}\left(t,x,y,\,z\right)\,=\,\frac{1}{2}N^{\mp}\left[ X_{i}^{R}\,e^{\pm\,\frac{i\pi}{2}\left(i\mathcal{M}-\frac{1}{2}\right)}\,\int_{-\infty}^{+\infty}e^{i\,\kappa\left[ x\,\sinh\vartheta\,\mp\,t\,\cosh\vartheta\right]}\,e^{\mp\left(i\mathcal{M}-\frac{1}{2}\right)\vartheta}\,d\vartheta\,+\right. \nonumber \\ +\left.Y_{i}^{R}\,e^{\pm\,\frac{i\pi}{2}\left(i\mathcal{M}+\frac{1}{2}\right)}\,\int_{-\infty}^{+\infty}e^{i\,\kappa\left[ x\,\sinh\vartheta\,\mp\,t\,\cosh\vartheta\right]}\,e^{\mp\left(i\mathcal{M}+\frac{1}{2}\right)\vartheta}\,d\vartheta \right]\times \nonumber \\ \times\,e^{i\,k_{2}\,y\,+\,i\,k_{3}\,z} \;\;\;\;\;\;\;\; \label{eq:minkbesmodsp}\eea
 These are the global functions we were looking for. For them, the following properties hold true:
 \begin{itemize}
 \item they are well behaved (analytical) on the entire Minkowski manifold, except for the origin;
 \item they are solutions of the Dirac equation;
 \item they are eigenfunctions of the boost generator operator $M_{01}$, with eigenvalue $\mathcal{M}$;
 \item they reduce themselves to the correct solutions of the Dirac equations in the different sectors;
 \item they correspond each to one of the two possible paths of analytical continuation we mentioned before, namely $\Psi_{i\mathcal{M}}^{-}$ corresponds to the path A, and $\Psi_{i\mathcal{M}}^{+}$ corresponds to the path B, so we could say that the reason for the existence of two different global representation is the existence of two different paths of analytical continuation across the horizons;
 \item they are orthonormalized with respect to the ordinary scalar product in MS, with normalization factors given by:
 \be N^{\mp}=\frac{e^{\pm\frac{1}{2}\pi\mathcal{M}}}{2\pi\sqrt{\kappa}} \ee 
 \end{itemize} 
 We note also that these functions represent the analogous in the
 spinor case of the Gerlach's Minkowski Bessel modes for the scalar
 field ~\cite{Gerlach}. Before considering the quantization of the
 field in terms of these modes, it is worth to notice that we could
 expect to have additional difficulties in using them instead of the
 standard plane wave basis. The reason for that is the divergence of
 these modes in the origin of Minkowski space, which does't affect, of
 course, the quantization procedure in any deep way, as will be proved
 in the following, but requires additional care in the calculations.

\subsection{Alternative quantization in MS}
 Having obtained these global functions, we can perform the quantization of the spinor field in terms of them. We remember that the usual plane wave expansion is given by:
 \be \Psi\,=\,\sum_{r\,=\,1,2}\int d^{3}k\,\left[a_{r}(k)\,\Psi_{r}^{+}(k)\,+\,b_{r}^{\dagger}(k)\,\Psi_{r}^{-}(k)\right] \label{eq:esppian} \ee  
 where the $\Psi_{r}^{+}(k)$ are positive frequency plane waves and $\Psi_{r}^{-}(k)$ are negative frequency ones, and that the quantum vacuum state is defined by the relation:
 \be a_{r}(k)\,\mid 0\,\rangle_{M}\,=\,b_{r}(k)\,\mid 0\,\rangle_{M}\,=\,0 \;\;\;\;\,\;\;\;\forall\,r\,,\,\vec{k} \ee

  But we know that our $\Psi_{i\mathcal{M}}^{-}$ are linear combinations of positive frequency plane waves and  $\Psi_{i\mathcal{M}}^{+}$ of negative frequency ones, so we can have this kind of expansion:
 \be \Psi(t,x,y,z)=\sum_{i=1,2}\int_{-\infty}^{+\infty}d\mathcal{M}\int dk_{2}\int dk_{3}\left[ c_{i\mathcal{M}}(k_{2},k_{3})\Psi_{i\mathcal{M}}^{-}+d^{\dagger}_{i\mathcal{M}}(k_{2},k_{3})\Psi_{i\mathcal{M}}^{+}\right] \label{eq:esplor} \ee 
 then imposing the usual anticommutations rules
 \be \left\{ c_{i\mathcal{M}}(k_{2},k_{3})\,,\,c^{\dagger}_{j\mathcal{M}'}(k'_{2},k'_{3}) \right\} \,=\,\delta_{ij}\,\delta\left( \mathcal{M}-\mathcal{M}'\right)\,\delta\left( k_{2}\,-\,k_{2}'\right)\delta\left( k_{3}\,-\,k_{3}'\right) \ldots\ldots \label{eq:regcomm}\ee
 so defining a vacuum state $\mid 0 \rangle$ by means of:
 \be c_{i\mathcal{M}}(k_{2},k_{3})\,\mid 0\,\rangle\,=\,d_{i\mathcal{M}}(k_{2},k_{3})\,\mid 0\,\rangle\,=\,0\,\;\;\,\;\;\;\,\;\;\;\forall\,i\,,\,\mathcal{M}\,,\,k_{2}\,,\,k_{3} \ee
 It is easy now to show that this quantization in equivalent to the usual one and so that the state $\mid 0 \rangle$ is the usual Minkowski vacuum state $\mid 0\rangle_{M}$.
 In fact we have the following relations:
\bea \lefteqn{c_{i\mathcal{M}}(k_{2},k_{3})\,=\,\left( \Psi_{i\mathcal{M}}^{-}\,,\,\Psi\right)_{M}\,=\,\int_{M}d^{3}x\,\Psi_{i\mathcal{M}}^{-\,\dagger}\,\Psi\,=} \nonumber \\ &=& \sum_{r}\,\int_{0}^{+\infty}dk'_{1}\,\frac{2\,\pi^{2}}{\omega'}\,N^{-}\,N_{k'}\,\left[  e^{i\,\frac{\pi}{2}\left( i\mathcal{M}+\frac{1}{2}\right)}\,e^{\left( i\mathcal{M}+\frac{1}{2}\right)\vartheta'}\,X_{i}^{\dagger}u_{r}(k')\,+ \nonumber \right. \\ &+&\left.\,e^{i\,\frac{\pi}{2}\left( i\mathcal{M}-\frac{1}{2}\right)}\,e^{\left( i\mathcal{M}-\frac{1}{2}\right)\vartheta'}\,Y_{i}^{\dagger}u_{r}(k')\right]\,\times\,a_{r}\left(k'_{1},k_{2},k_{3}\right)\,= \label{eq:c}\\ &=&\,\sum_{r}\,\int_{0}^{+\infty}dk'_{1}\,F_{i\,r}\left( k'_{1}\,,\,\mathcal{M}\right)\,a_{r}\left(k'_{1},k_{2},k_{3}\right) \eea
 with $\vartheta'=\frac{1}{2}\,\ln\left(\frac{\omega'\,+\,k'_{1}}{\omega'\,-\,k'_{1}}\right)$

  and, for $d_{i\mathcal{M}}^{\dagger}(k_{2},k_{3})$:
 \bea \lefteqn{d_{i\mathcal{M}}^{\dagger}(k_{2},k_{3})\,=\,\left( \Psi_{i\mathcal{M}}^{+}\,,\,\Psi\right)\,=\,\int_{M}d^{3}x\,\Psi_{i\mathcal{M}}^{+\,\dagger}\,\Psi\,=} \nonumber \\ &=& \sum_{r}\,\int_{0}^{+\infty}dk'_{1}\,\frac{2\,\pi^{2}}{\omega'}\,N^{+}\,N_{k'}\,\left[  e^{i\,\frac{\pi}{2}\left( i\mathcal{M}+\frac{1}{2}\right)}\,e^{\left( i\mathcal{M}+\frac{1}{2}\right)\vartheta'}\,X_{i}^{\dagger}v_{r}(k')\,+ \right. \nonumber \\ &+& \left.\,e^{i\,\frac{\pi}{2}\left( i\mathcal{M}-\frac{1}{2}\right)}\,e^{\left( i\mathcal{M}-\frac{1}{2}\right)\vartheta'}\,Y_{i}^{\dagger}v_{r}(k')\right]\,\times\,b^{\dagger}_{r}\left(k'_{1},k_{2},k_{3}\right)\,= \\ &=&\,\sum_{r}\,\int_{0}^{+\infty}dk'_{1}\,G_{i\,r}\left( k'_{1}\,,\,\mathcal{M}\right)\,b^{\dagger}_{r}\left(k'_{1},k_{2},k_{3}\right) \eea

  So it is demonstrated that the vacuum states defined by the two quantization procedures are the same. Moreover, by explicit calculation it is possible to show that
 \bea \lefteqn{\left\{ a_{s}\left( k''_{1},k_{2},k_{3}\right)\,,\,a_{r}^{\dagger}\left( k'_{1},k'_{2},k'_{3}\right)\right\}\,=\,\delta_{rs}\,\delta\left(k'_{1}-k''_{1}\right)\,\delta\left( k_{2}-k'_{2}\right)\,\delta\left(k_{3}-k'_{3}\right)\;\;\;\Longleftrightarrow} \nonumber \\ &\Longleftrightarrow&\left\{ c_{i\mathcal{M}}(k_{2},k_{3}),c^{\dagger}_{j\mathcal{M}'}(k'_{2},k'_{3})\right\}=Cost\times\delta_{ij}\delta\left(\mathcal{M}-\mathcal{M}'\right)\delta\left( k_{2}-k'_{2}\right)\delta\left(k_{3}-k'_{3}\right) \nonumber \eea
 so the two quantum constructions are totally equivalent.

% This means also that the divergence of the boost modes at the origin
% of MS does not create any qualitative difficulty for the
% quantization procedure, but just requires some additional care in
% the calculations we will perform in the following.

\subsection{The Unruh construction and the Unruh effect} \label{sec:Un}
 We will now derive the Unruh effect for the spinor field following
the standard procedure first used by Unruh himself.
% The main idea is to have a quantum construction which is valid for MS
%and reproduces the construction outlined in ~\ref{sec:RQFT} for the R sector. 

Let's first recall that an attempt in finding the relationship between
the quantum construction in RS and that in MS was made by Fulling
~\cite{Fulling} (for the scalar field), who simply identified the RS
with the R-sector of the MS, consequently considered the Rindler
vacuum state as a state in the Minkowski Hilbert space, and tried to
express the Rindler annihilation (and creation) operators in terms of
the usual plane waves ones. He then argued that Minkowski vacuum state
could be considered as a particle state with respect to the Rindler
vacuum state. Of course, because of the boundary condition we found
necessary for the quantization in RS, this procedure is meaningless,
since, as we already stressed, RS cannot be identified with R-sector
of MS, but should be considered as a manifold on its own. Apart from
this, however, there is another reason why the Fulling scheme is not
valid. This scheme, in fact, implies to consider field modes in MS which
corresponds to the Fulling ones in R and are zero everywhere
else. This is equivalent to use a representation of the boost modes
~\ref{eq:minkbesmodsp} given by

\bea
\Psi_{\mathcal{M}}\,=\,\theta(x_{+})\theta(-x_{-})\Psi_{\mathcal{M}}^{R}\,+\,\theta(x_{+})\theta(x_{-})\Psi_{\mathcal{M}}^{F}\,+\,\theta(-x_{+})\theta(x_{-})\Psi_{\mathcal{M}}^{L}\,+\nonumber
\\ +\,\theta(-x_{+})\theta(-x_{-})\Psi_{\mathcal{M}}^{P}
\eea
with $x_{\pm}=t\pm x$, but then using for the quantization only the
first term of this expression, so to be restricted in the R-wedge. This
procedure cannot be valid, since physically these modes
correspond to solutions of the field equation when infinite sources of
energy are placed at the horizon, because of the presence of the theta
function.

The procedure used by Unruh to compare the Minkowski quantization to
the Rindler one is to construct a new quantization scheme, which
should be valid in the whole MS, but should also reproduce the Fulling
quantization in RS when restricted to to the R-sector. The idea is to
build the Hilbert space of Minkowski states $\mathcal{H}_{M}$ out of
the Hilbert space of solutions of the wave equation of the form
$\mathcal{H}_{R}\oplus\mathcal{H}_{L}$, i.e.  sum of solution of
Minkowski equation of motion which are the same as the Fulling modes
in the R (L) sector, but vanish identically in the L (R) sector.

Let's perform the Unruh construction for quantization in MS, using our globally defined functions $\Psi_{i\mathcal{M}}^{\mp}$.
Consider the functions:
\bea
\lefteqn{R_{i\mathcal{M}}\,=\,\frac{1}{\sqrt{2\,\cosh\pi\mathcal{M}}}\,\left(
e^{\frac{\pi\,\mathcal{M}}{2}}\,\Psi_{i\mathcal{M}}^{-}\,+\,e^{-\,\frac{\pi\,\mathcal{M}}{2}}\,\Psi_{i\mathcal{M}}^{+}\right)}\label{eq:RU}
\\
&L_{i\mathcal{M}}&\,=\,\frac{1}{\sqrt{2\,\cosh\pi\mathcal{M}}}\,\left(
e^{-\,\frac{\pi\,\mathcal{M}}{2}}\,\Psi_{i\mathcal{M}}^{-}\,-\,e^{\frac{\pi\,\mathcal{M}}{2}}\,\Psi_{i\mathcal{M}}^{+}\right) \label{eq:LU}\eea
which are solutions of Dirac equations, are eigenfunctions of
$M_{01}$, are analytical in the whole Minkowski space, except fro the origin, and orthonormalzed in MS. Moreover, it happens that the $R_{i\mathcal{M}}$ are defined everywhere but in L sector and reduce themselves to the $\Psi_{i\mathcal{M}}^{R}$ in the R one, while the $L_{i\mathcal{M}}$ manifest the inverse behaviour. Inverting these relations, and inserting into the expansion ~\ref{eq:esplor}, we have:
\bea \lefteqn{\Psi(t,x,y,z)=\sum_{i\,=\,1,2}\int_{-\infty}^{+\infty} d\mathcal{M}\int dk_{2}\int dk_{3}\left[ c_{i,\mathcal{M}}(k_{2},k_{3})\Psi_{i,\mathcal{M}}^{-}+d^{\dagger}_{i,\mathcal{M}}(k_{2},k_{3})\Psi_{i,\mathcal{M}}^{+}\right]\,=} \nonumber \\  &=& \sum_{i}\int_{-\infty}^{+\infty} d\mathcal{M}\left[ r_{i,\mathcal{M}}\,R_{i,\mathcal{M}}\,+\,l^{\dagger}_{i,\mathcal{M}}\,L_{i,\mathcal{M}}\right]\,=\\ &=& \sum_{i}\int_{0}^{+\infty} d\mathcal{M}\left[ r_{i,\mathcal{M}}\,R_{i,\mathcal{M}}\,+\,l_{i,\mathcal{M}}\,L_{i,-\mathcal{M}}\,+\,r^{\dagger}_{i,\mathcal{M}}\,R_{i,-\mathcal{M}}\,+\,l^{\dagger}_{i,\mathcal{M}}\,L_{i,\mathcal{M}}\right]\;\;\;\;\;\,\;\;\;\;\;\; \label{eq:espUnr} \eea
having introduced the coefficients:
\bea r_{i,\mathcal{M}}\,=\,\frac{c_{i,\mathcal{M}}\,e^{\frac{\pi\mathcal{M}}{2}}\,+\,d^{\dagger}_{i,\mathcal{M}}\,e^{-\frac{\pi\mathcal{M}}{2}}}{\sqrt{2\,\cosh\pi\mathcal{M}}} \label{eq:Bog} \\ l^{\dagger}_{i,\mathcal{M}}\,=\,\frac{c_{i,\mathcal{M}}\,e^{-\frac{\pi\mathcal{M}}{2}}\,-\,d^{\dagger}_{i,\mathcal{M}}\,e^{\frac{\pi\mathcal{M}}{2}}}{\sqrt{2\,\cosh\pi\mathcal{M}}} \eea
These relations represent the Bogolubov transformation between the quantum constructions (~\ref{eq:esplor}) and (~\ref{eq:espUnr}).

Now, consider carefully the nature of the operators
$r_{i,\mathcal{M}}$ and $l_{i,\mathcal{M}}$ (and
$r^{\dagger}_{i,\mathcal{M}}$ and
$l^{\dagger}_{i,\mathcal{M}}$). 

First of all suppose that the (~\ref{eq:espUnr}) represents an
expansion of the spinor field, that leads to a correct quantization of it in MS, providing we impose the conditions

\be
\left\{r_{i,\mathcal{M}},r^{\dagger}_{j,\mathcal{M}'}\right\}\,=\,\delta_{ij}\delta(\mathcal{M}-\mathcal{M}')
\;\;\;\;\left\{
r_{i,\mathcal{M}},r_{j,\mathcal{M}}\right\}\,=\,0\;\;\;\; \left\{
r_{i,\mathcal{M}},l_{j,\mathcal{M}}\right\}\,=\,0 \label{eq:ant} \ee
 and the analogous for $l_{i,\mathcal{M}}$ and
$l^{\dagger}_{i,\mathcal{M}}$.
   
Suppose also that the operators  above could be considered as  annihilation
(and creation) operators for R-particles and L-particles.

We stress that these hypothesis are crucial for the following derivation being physically meaningful.

In fact, if these both hold, the operators $r_{i,\mathcal{M}}$, defined as
scalar products in MS, could be also expressed as integrals over the
surface ($t=0$, $x > 0$) which is a Cauchy (hyper)surface for the R
wedge.

Moreover, this, together with the particular functional behaviour of
the R-function (that reduce to the $\Psi_{i\mathcal{M}}^{R}$ in the R
sector), would mean that we can identify the operators
$r_{i,\mathcal{M}}$ ($r^{\dagger}_{i,\mathcal{M}}$) with the Rindler
annihilation (creation) operators $a_{i,\mathcal{M}}$
($a^{\dagger}_{i,\mathcal{M}}$). Consequently, the R-particles would
be identified with the Rindler particles, constructed in terms of the
$a^{\dagger}_{i,\mathcal{M}}$, that an accelerated observer detects.   

Let's now consider the operator $N_{i,j,\mathcal{M},\mathcal{M}'}=r_{i\mathcal{M}}^{\dagger}r_{j\mathcal{M}'}$ (we can of course not consider the quantum numbers $k_{2}$ and $k_{3}$, because they don't play any significant role here). We remind that $\mathcal{M}$ is the eigenvalue of $M_{01}$ but also the energy of a Rindler observer. It is easy to calculate the mean value of this operator in the Minkowski vacuum state, having:
\bea \lefteqn{_{M}\langle 0\,\mid\,N_{i,j,\mathcal{M},\mathcal{M}'}\,\mid 0\,\rangle_{M}\,=\,_{M}\langle 0\,\mid\,r_{j\mathcal{M}'}^{\dagger}r_{i\mathcal{M}}\,\mid 0\,\rangle_{M}\,=} \\  &=& \frac{1}{e^{2\pi\mathcal{M}}\,+\,1}\,\delta_{ij}\,\delta\left(\mathcal{M}-\mathcal{M}'\right)\,\delta\left( k_{2}-k'_{2}\right)\delta\left( k_{3}-k'_{3}\right) \label{eq:numpart}\eea

If we now identify the operators $r_{i,\mathcal{M}}$ with the
operators $a_{i,\mathcal{M}}$, then we can intepret
$N_{i,j,\mathcal{M},\mathcal{M}'}$ as a Rindler particle number
operator. This is exactly what is usually done in literature. The
reasons for this identification were explained before and appear to be
quite convincing.  Nevertheless this passage is not trivial at all, as we will show.

Anyway, once identified the operators $r_{i,\mathcal{M}}$ with the operators $a_{i,\mathcal{M}}$, the result (~\ref{eq:numpart}) can be interpreted as meaning that an inertial observer and a Rindler observer don't share the same vacuum state, and that Minkowski vacuum state is a particle state for a Rindler observer.

We can also calculate the total number of particles that a Rindler observer will perceive if the field is in Minkowski vacuum state, for any quantum number and for unity of proper time.
The result is:
\bea \;\;\; \lefteqn{dN\,=\,\sum_{i,j}\int_{0}^{+\infty}d\mathcal{M}\int_{0}^{+\infty}d\mathcal{M}'\int dk_{2}\int dk'_{2}\int dk_{3}\int dk'_{3}\,N_{i,j,\mathcal{M},\mathcal{M}'}\,d\eta\,=} \\ &=&\,\sum_{i,j}\int_{0}^{+\infty}d\mathcal{M}\int_{0}^{+\infty}d\mathcal{M}'\int dk_{2}\int dk'_{2}\int dk_{3}\int dk'_{3} \,_{M}\langle 0\,\mid\,r_{j\mathcal{M}'}^{\dagger}r_{i\mathcal{M}}\,\mid 0\,\rangle_{M}\,d\eta\,= \nonumber \\ &=& \sum_{i,j}\int_{0}^{+\infty}d\mathcal{M}\int_{0}^{+\infty}d\mathcal{M}'\int dk_{2}\int dk'_{2}\int dk_{3}\int dk'_{3}\,\frac{1}{e^{2\pi\mathcal{M}}\,+\,1}\,\times \\ &\times& \delta_{ij}\,\delta\left(\mathcal{M}-\mathcal{M}'\right)\,\delta\left( k_{2}-k'_{2}\right)\delta\left( k_{3}-k'_{3}\right)\,d\eta\,= \\ &=& \sum_{i}\int_{0}^{+\infty}d\mathcal{M}\int dk_{2}\int dk_{3}\,\frac{1}{e^{2\pi\mathcal{M}}\,+\,1}\,d\eta\,= \\ &=&\,\sum_{i}\int_{0}^{+\infty}dh_{R}\int dk_{2}\int dk_{3}\,\frac{1}{e^{\frac{2\pi\,h_{R}}{a}}\,+\,1}\,d\tau\,= \\ &=& 2\,\int_{0}^{+\infty}dh_{R}\int dk_{2}\int dk_{3}\,\frac{1}{e^{\frac{2\pi\,h_{R}}{a}}\,+\,1}\,d\tau   \label{eq:flussoU} \eea
where we have used the quantities: $h_{R}=a\mathcal{M}$, Rindler energy, and $\tau=\frac{\eta}{a}$, proper time, with $a$ acceleration of the Rindler observer.
This result could be stated saying that the particle distribution of the Minkowski vacuum state, with respect to the quantization performed using the $R_{i\mathcal{M}}$ modes, is given by a thermal spectrum, according to Fermi-Dirac statistics with temperature:
\be T\,=\,\frac{1}{\beta\,k_{B}}\,=\,\frac{a}{2\,\pi\,k_{B}} \label{eq:TTT}\ee

We saw that the identification between the operators
$r_{i,\mathcal{M}}$ and $a_{i,\mathcal{M}}$ is crucial in its
interpretation. We also pointed out that this identification is based
on two strong hypotheses: that the Unruh construction is valid in the
whole MS, and that the operators $r_{i,\mathcal{M}}$
($r^{\dagger}_{i,\mathcal{M}}$), and  $l_{i,\mathcal{M}}$
($l^{\dagger}_{i,\mathcal{M}}$) can be considered as annhilation
(creation) operators.

Consequently, we will now analyse in detail these crucial points.

\subsection{Analysis of the Unruh construction}
We are going now to show that the same arguments against the validity of the
Unruh construction in the whole Minkowski Space, that were indicated
in ~\cite{Bel1}~\cite{Bel2}~\cite{Bel3}~\cite{Bel4} for the case of
the scalar field, are preserved also for the spinor field and support the
conclusion that the Unruh scheme  is valid instead in the disjont union of the R
and L wedge. This conclusion will be also proved by an explicit
calculation.  

First of all we argue that the Unruh quantization scheme is not
suitable for the quantization of the spinor field in MS. This can be 
easily  seen by looking at the initial expansion of the field
(~\ref{eq:espUnr}). This is based on a separation of the integration
interval into two parts corresponding to positive and negative values
of the eigenvalue $\mathcal{M}$, and this separation is necessary in
order to have an expansion in terms of R and L modes. But we should
recall the divergence of the functions $\Psi_{i,\mathcal{M}}$ (hence
of the Unruh R and L modes) at the
origin of MS, as we already noticed. This, we said, doesn't affect the
quantization in MS in terms of the boost modes, but nevertheless implies
that we cannot perform the separation of the integration interval as
in the Unruh procedure. In fact, discarding the eigenfunction
correspondent to the eigenvalue $\mathcal{M}=0$ (as any other
eigenfunction, because the divergence at the origin is a common
property of all of them) it would mean to
discard an infinite number of degrees of freedom of the field, if the
origin is in the domain of definition of our field, so
commuting from our initial system to a phisically different
one. Consequently we could say that the Unruh construction cannot be
valid in the whole MS, which of course include the origin.

Moreover, the Unruh operators $r$ and $l$ (and their conjugates)
cannot be considered as annihilation (and creation) operators for the
field in MS, even if they satisfy the anticommutation relations
(~\ref{eq:ant}). For this being possible, it is necessary the existence
of a stationary ground state for the field in MS, which is defined
with respect to a global timelike variable in the whole space, and
which is annihilated by the $r$ and $l$ operators. But such a ground
state is definitely missing in MS. In fact, there is no timelike
variable with respect to which the Unruh modes (or their adjoints) are
positive frequency solutions of the Dirac equation in MS (remember
that the Killing vector $\frac{\partial}{\partial\eta}$ is spacelike
in the $F$ and $P$ sectors) and consequently the Unruh operators will
always be linear combinations of operators which create or annihilate
particles with opposite frequency signs, i.e. there is not a
stationary vacuum state in MS with respect to r-particles or
l-particles.

These problems disappear if we consider the field only defined in the
double Rindler wedge, i.e. in the disjoint union of the R and L
sectors. Here the origin of MS is not considered, and so the divergence of the $R$ and $L$
modes has no physical consequencies for the separation of the
integration domain in the expansion (~\ref{eq:espUnr}). In this
manifold there exists a timelike killing vector with respect to which
the $R$ and $L$ modes are positive frequency solutions of the Dirac
equation, consequently there exists a stationary ground state for the
system defined by the Unruh expansion, and the $r$ and $l$ operators
could be interpreted as annihilation operators for the field in this
double wedge. The result is that the Unruh construction is well
defined in this case but, we emphasize, it refers to a field
restricted to the union of the R and L sectors of MS and so physically
different from the field defined in MS. In addition, since the R and L
sectors are totally independent of each other, because they are
separated by a spacelike interval, the quantization in these two
wedges should be carried on separately. In other words we have the
expansion:
\bea
\Psi_{RL}(x)\,=\,\sum_{i}\int_{0}^{\infty}d\mathcal{M}\left\{\,r_{i,\mathcal{M}}R_{i,\mathcal{M}}(x)\,+\,r^{\dagger}_{i,\mathcal{M}}R_{i,-\mathcal{M}}(x)\right\}\,+\nonumber
\\ +\,\sum_{i}\int_{0}^{\infty}d\mathcal{M}\left\{\,l_{i,\mathcal{M}}L_{i,-\mathcal{M}}(x)\,+\,l^{\dagger}_{i,\mathcal{M}}L_{i,\mathcal{M}}(x)\right\}\;\;\;\;\;\;x\,\in\,R\,\cup\,L
\eea

We already showed that the quantization of the spinor field in the R sector requires the boundary condition (~\ref{eq:Condizione}), so we
expect it to be manifest also for the field $\Psi_{RL}$. In particular,
if the Unruh espansion is valid (only) for the field in the double
Rindler wedge, the Unruh operators $r$ should coincide with
the operators $a_{i,\mathcal{M}}$ in terms of which the quantization
of the field in RS is performed, provided that the field satisfy the
boundary condition (~\ref{eq:Condizione}). Let's now prove with an explicit
calculation this statement.  

First of all, we are going to find the explicit expression of the coefficients $r_{i\mathcal{M}}$ as functions of the values of the field $\Psi_{RL}$.
We recall that these are defined as:
\be r_{i,\mathcal{M}}\,=\,\frac{c_{i,\mathcal{M}}\,e^{\frac{\pi\mathcal{M}}{2}}\,+\,d^{\dagger}_{i,\mathcal{M}}\,e^{-\frac{\pi\mathcal{M}}{2}}}{\sqrt{2\,\cosh\pi\mathcal{M}}} \label{eq:r}\ee 
so we first need to find the expression for the coefficients $c_{i,\mathcal{M}}$  and $d^{\dagger}_{i,\mathcal{M}}$, in terms of the field and of its spatial derivative. This task implies the proper treatment of integral whose hypersurface of integration is the $t=0$ hypersurface, which pass across the origin of Minkowski space, that is the intersection of the branch points where the functions we used are not well defined; this requires great attention.

By performing this calculation it is also possible to see that, as it should be, the coefficients $c_{i,\mathcal{M}}$  and $d^{\dagger}_{i,\mathcal{M}}$ are well defined without the need of any additional boundary condition other than the vanishing of the field at spatial infinity, in contrast to the \lq\lq Rindler coefficients" in (~\ref{eq:exp}).

Inserting the expressions so found for $c_{i,\mathcal{M}}$  and $d^{\dagger}_{i,\mathcal{M}}$ into the (~\ref{eq:r}), we obtain:
\bea \lefteqn{r_{i,\mathcal{M}}\,=} \nonumber \\ &=&
\frac{1}{2\,\pi\,\sqrt{\kappa}}\,\sqrt{2\,\cosh\pi\mathcal{M}}\,\int_{-\infty}^{+\infty}dy\,e^{-ik_{2}y}\,\int_{-\infty}^{+\infty}dz\,e^{-ik_{3}z}\,\times
\nonumber \\ &\times& \left\{
X_{i}^{\dagger}\,\int_{0}^{+\infty}dx\,\left[
K_{i\mathcal{M}+\frac{1}{2}}(\kappa x)\Psi_{RL}\left(
0\,,\,x\right)\,-\right.\right. \nonumber \\
&-&\,\frac{1}{2}\,\Gamma\left( \frac{1}{2}+
i\mathcal{M}\right)\left(\frac{\kappa x}{2}\right)^{-
i\mathcal{M}-\frac{1}{2}}\,\Psi_{RL}\left(0\,,\,x\right)\,- \nonumber \\
&-&\left.\,\frac{1}{\kappa}\,\frac{1}{\frac{1}{2}-
i\mathcal{M}}\,\Gamma\left( \frac{1}{2}+
i\mathcal{M}\right)\left(\frac{\kappa x}{2}\right)^{-
i\mathcal{M}+\frac{1}{2}}\,\frac{d}{dx}\Psi\left(0\,,\,x\right)\right]\,+
\nonumber \\ &+& Y_{i}^{\dagger}\,\int_{0}^{+\infty}dx\,\left[
K_{i\mathcal{M}-\frac{1}{2}}(\kappa x)\Psi_{RL}\left(
0\,,\,x\right)\,-\right. \nonumber \\ &-&\,\frac{1}{2}\,\Gamma\left(
\frac{1}{2}- i\mathcal{M}\right)\left(\frac{\kappa x}{2}\right)^{+
i\mathcal{M}-\frac{1}{2}}\,\Psi_{RL}\left(0\,,\,x\right)\,- \nonumber \\
&-&\left.\left.\,\frac{1}{\kappa}\,\frac{1}{\frac{1}{2}+
i\mathcal{M}}\,\Gamma\left( \frac{1}{2}-
i\mathcal{M}\right)\left(\frac{\kappa x}{2}\right)^{+
i\mathcal{M}+\frac{1}{2}}\,\frac{d}{dx}\Psi_{RL}\left(0\,,\,x\right)\right]
\right\} \label{eq:r1}\eea
Changing coordinates to the Rindler ones (R sector), we have:
\bea \lefteqn{r_{i,\mathcal{M}}\,=} \nonumber \\ &=& \sqrt{2}\,a_{i,\mathcal{M}}\,+\,\frac{1}{2\,\pi\,\sqrt{\kappa}}\,\sqrt{2\,\cosh\pi\mathcal{M}}\,\int_{-\infty}^{+\infty}dy\,e^{-ik_{2}y}\,\int_{-\infty}^{+\infty}dz\,e^{-ik_{3}z}\,\times \nonumber \\ &\times& \left\{ \lim_{\rho\rightarrow 0}\,\frac{1}{2}\,\frac{1}{\frac{1}{2}- i\mathcal{M}}\,\Gamma\left( \frac{1}{2}+ i\mathcal{M}\right)\left(\frac{\kappa \rho}{2}\right)^{- i\mathcal{M}+\frac{1}{2}}\,X_{i}^{\dagger}\,\Psi^{R}\left(0\,,\,\rho\right)\,+\right. \nonumber \\ &+&\,\left.\lim_{\rho\rightarrow 0}\,\frac{1}{2}\,\frac{1}{\frac{1}{2}+ i\mathcal{M}}\,\Gamma\left( \frac{1}{2}- i\mathcal{M}\right)\left(\frac{\kappa \rho}{2}\right)^{+ i\mathcal{M}+\frac{1}{2}}\,Y_{i}^{\dagger}\,\Psi^{R}\left(0\,,\,\rho\right)\right\} \eea

It is so clear that the coefficients $r_{i,\mathcal{M}}$ and $a_{i,\mathcal{M}}$ cannot be identified unless
\be \lim_{\rho\,\rightarrow\,0}\,\rho^{\frac{1}{2}}\,\Psi^{R}\left( 0\,,\,\rho\right)\,=\,0 \label{eq:co} \ee
which is just the boundary condition we found in sec.~\ref{sec:RQFT}.

Let's now discuss this result.
%\subsection{Discussion of the result}
%\begin{itemize}
%\item 
We saw that it is not possible to identify the operators $r_{i,\mathcal{M}}$ and $a_{i,\mathcal{M}}$, unless (~\ref{eq:co}); but
%\item 
there are no physical reasons to impose this boundary condition on the spinor field in MS;  
%\item this would not be an operation without serious outcoming, because it wou%ld dramatically change the properties of MS: it would 
%be no more a globally hyperbolic spacetime and the hypersurface $t=0$ (on whic%h $r_{i,\mathcal{M}}$ themselves are defined) would %not be a Cauchy surface a%nymore;
%\item 
%this problem could be avoided 
we have seen indeed that we obtain a meaningful quantum theory in
terms of the Unruh modes only if we perform the \lq\lq Unruh construction" just in the R and L sectors of MS, which are completely disjoint, from the causal and physical point of view;
moreover, for an observer living in the Rindler wedge it is not possible, because of condition (~\ref{eq:co}), to perform measurement in the whole MS, in order to put the field in the Minkowski vacuum state;  
%\item 
in other words, the Unruh construction outlined in \ref{sec:Un} does
not represent a valid quantization scheme for the whole MS;
consequently the operator $r^{\dagger}r$ cannot be interpreted as a
particle number operator in MS and therefore, the relation (~\ref{eq:flussoU}) cannot be interpreted in any sense as a proof of the \lq\lq Unruh effect".
%\end{itemize}

\section{Conclusions}
Here we come to our conclusions. We can say that our analysis of the
spinor field confirms the conclusion made in
~\cite{Bel1}~\cite{Bel2}~\cite{Bel3}~\cite{Bel4} that the basic
principles of quantum field theory imply that the Unruh procedure does not represent a correct derivation of the Unruh effect.

We saw that the reason for this conclusion is the existence of
boundary conditions for the quantization of the spinor field in
Rindler spacetime, preventing any relationship between this quantum
construction and that one in Minkowski spacetime.The role played by
this boundary conditions was analytically showed in our analysis of the Unruh procedure. We already noted that the existence of such boundary conditions could be expected since a Rinlder observer is confined inside the Rindler wedge by event horizons, so he would see these like spatial infinity of an inertial observer.
For the same reason a Rindler observer has no relationship with MS and he cannot in any way prepare the quantum field in the Minkowski vacuum state.

We have also shown that the Unruh quantization scheme is valid only in
 the double Rindler wedge $R\cup L$, and so it is also in the spinor
 case, because of the boundary condition (~\ref{eq:Condizione});
 consequently the Unruh construction cannot be used to analyse the
 relationship between Rindler and Minkowski quantization schemes. 

 It seems to us that there are enough reasons to assert that this is a general
feature of the analysed problem and that it holds true for any quantum
field, and we are supported in this conclusion by the previously
mentioned and similar results obtained for the scalar field and for
the electromagnetic field.

On the other hand, the appearence of the fermion factor in the
distribution (~\ref{eq:flussoU}) is entirely due to the particular
form of the Bogolubov transformation (~\ref{eq:Bog}), and there is no
real need to interpretate it as an proof of a thermal nature of the
spectrum; a similar situation is encounterd in other physical problems
where the concept of temperature doesn't arise at all, as it is explained in details in ~\cite{Bel3}.

Of course, there are many different approaches to the Unruh problem and many derivations of the effect have been proposed. It is then worth to study these carefully in order to see if the difficulties found here for the Unruh procedure survive also in these cases, or they can be considered as correct derivations of the Unruh effect and of its consequences. Particularly interesting are the results in \cite{TVD} and \cite{Pare}, which deserve further study and attention.

As regards to the other aspect of the Unruh problem, that we mentioned in the beginning, namely the behaviour of an accelerated detector in Minkowski space, we could only say that it remains an open question. This should be clear also just considering that the Unruh effect is generally explained using the key role of the event horizons, but there are no horizon at all for a non-ideal accelerating detector, whose acceleration lasts a finite amount of time. 

Our results show, however, that there are no reasons to expect that it will be of the type predicted by Unruh, at least as far as only the conventional derivation of it is considered, and that it should not be expected to be universal and independent from the nature and characteristics of the detector itself. It is worth to note that, as a partial confirmation of what we are saying, it was showed in great details in ~\cite{NikRit} that elementary particles accelerated by a constant electric field don't follow, in general, the Unruh behaviour, i.e. the thermal response with temperature (\ref{eq:T}).

We are sure, of course, that an accelerated detector behaves, in general, in a different way from an inertial one, and we admit that in some cases it could follow the Unruh behaviour, but we think that there are no quantum theoretical reasons to believe that this is the universal one. 
\section*{Acknowledgements} 
We would like to thank here V. A. Belinski, for enlightening
discussions and advices, and for having read the manuscript of this paper, giving
useful comments. A special thank goes also to B.  Narozhny, who
noticed an error occurred in a previous version of this work, now
corrected, as well as U. Gerlach for useful comments.

\end{document}